\documentclass[a4paper,fleqn]{cas-dc}

\usepackage[numbers,sort&compress]{natbib}
\usepackage{siunitx}
\usepackage{microtype}

\def\tsc#1{\csdef{#1}{\textsc{\lowercase{#1}}\xspace}}
\tsc{WGM}
\tsc{QE}
\tsc{EP}
\tsc{PMS}
\tsc{BEC}
\tsc{DE}
\begin{document}
\let\WriteBookmarks\relax
\def\floatpagepagefraction{1}
\def\textpagefraction{.001}

% Short title
\shorttitle{Structure constraint neural network}

% Short author
\shortauthors{Xiao jiang et~al.}

% Main title of the paper
\title [mode = title]{Towards interpreting the thermally activated $\beta$ dynamics in metallic glass with the structural constraint neural network}                      
% Title footnote mark
% eg: \tnotemark[1]
% \tnotemark[1,2]

% Title footnote 1.
% eg: \tnotetext[1]{Title footnote text}
% \tnotetext[<tnote number>]{<tnote text>} 
% \tnotetext[1]{This document is the results of the research
   % project funded by the National Science Foundation.}

% \tnotetext[2]{The second title footnote which is a longer text matter
   % to fill through the whole text width and overflow into
   % another line in the footnotes area of the first page.}

% First author
%
% Options: Use if required
% eg: \author[1,3]{Author Name}[type=editor,
%       style=chinese,
%       auid=000,
%       bioid=1,
%       prefix=Sir,
%       orcid=0000-0000-0000-0000,
%       facebook=<facebook id>,
%       twitter=<twitter id>,
%       linkedin=<linkedin id>,
%       gplus=<gplus id>]
\author[1]{Xiao Jinag}[]

%  Credit authorship
% \credit{Conceptualization of this study, Methodology, Software}

% Address/affiliation
\affiliation[1]{organization={College of Computer Science and Electronic Engineering,},
    addressline={Hunan University}, 
    city={Changsha},
    % citysep={}, % Uncomment if no comma needed between city and postcode
    postcode={410012}, 
    % state={},
    country={China}}

% Second author
\author[1]{Zean Tian}[]
% Corresponding author indication
\cormark[1]

% Footnote of the first author
% \fnmark[]

% Email id of the first author
\ead{tianzean@hnu.edu.cn}

% % URL of the first author
% \ead[url]{www.cvr.cc, cvr@sayahna.org}
% Third author
\author[1]{Kenli Li}[%
   ]
\author[2]{Wangyu Hu}[%
   ]

% \credit{Data curation, Writing - Original draft preparation}

% Address/affiliation
\affiliation[2]{organization={College of Materials Science and Engineering},
    % addressline={}, 
    city={Changsha},
    % citysep={}, % Uncomment if no comma needed between city and postcode
    postcode={410012}, 
    % state={Trivandrum},
    country={China}}

% Corresponding author text
\cortext[cor1]{Corresponding author. }

% Here goes the abstract
\begin{abstract}
Unraveling the structural factors influencing the dynamics of amorphous solids is crucial. While deep learning aids in navigating these complexities, transparency issues persist. Inspired by the successful application of prototype neural networks in the field of image analysis, this study introduces a new machine-learning approach to tackle the interpretability challenges faced in glassy research. Distinguishing from traditional machine learning models that only predict dynamics from the structural input, the adapted neural network additionally tries to learn structural prototypes under various dynamic patterns in the training phase. Such learned structural constraints can serve as a breakthrough in explaining how structural differences impact dynamics. We further use the proposed model to explore the correlation between the local structure and activation energy in the CuZr metallic glass. Building upon this interpretable model, we demonstrated significant structural differences among particles with distinct activation energies. The insights gained from this analysis serve as a data-driven solution for unraveling the origins of the structural heterogeneity in amorphous alloys, offering a valuable contribution to the understanding the amorphous materials.

\end{abstract}

% Use if graphical abstract is present
% \begin{graphicalabstract}
% \includegraphics{figs/grabs.pdf}
% \end{graphicalabstract}

% Research highlights
% \begin{highlights}
% \item This research overcomes the limitations of conventional machine learning approaches, providing a unique interpretable model for uncovering the intricate relationship between structure and dynamics.
% \item The model demonstrates commendable predictive accuracy in forecasting activation energies of CuZr alloys. It allows us to identify and understand the primary structural factors responsible for the differences in dynamics.
% \item The insights gained from the model serve as a data-driven solution for unraveling the origins of the structural heterogeneity in amorphous alloys.

% \end{highlights}

% Keywords
% Each keyword is seperated by \sep

\maketitle

\section{Introduction}

Amorphous alloys, also referred to as metallic glasses (MGs), represent a fascinating class of disordered materials\cite{greer1995metallic,cheng2011atomic,weeks2000three,yang2021determining}, characterized by their distinctive local atomic arrangements. Unlike crystalline materials, where structural defects are easily discernible due to the periodic arrangement of atoms in the lattice, metallic glasses lack both translational and rotational periodicity, posing a more intricate challenge for defect detection. Nevertheless, unraveling the mysteries of structural defects in amorphous solids is crucial for comprehending their complex physical behaviors \cite{speck2012first, marin2019slowing,berthier2007structure}. Parameters of physical intuition, derived from geometric or non-geometric parameters \cite{kawasaki2007correlation,hu2015five, tong2019structural,richard2020predicting}, offer insights into the connection between structural defects and physical properties. Despite these advancements, the origin of the underlying structure remains elusive.

Recently, machine learning techniques have emerged as a promising avenue to discern the structural aspects influencing the dynamical property. In the past five years, both supervised and unsupervised algorithms have been extensively applied to predict dynamic properties in amorphous alloys, covering aspects such as athermal local structural deformation \cite{wang2019transferable, yang2021machine,xu2023identification}, activation energy\cite{wang2020predicting}, long-time diffusion \cite{wu2023machine,wu2023unsupervised}, and so on \cite{peng2021machine,liu2022machine,liu2023concurrent,tao2023structural}. The model takes the geometric features as input, such as radial symmetry functions, interstice distribution, or solely particle position information \cite{cubuk2015identifying,cubuk2016structural, bapst2020unveiling, wang2021inverse,alkemade2022comparing,pezzicoli2022se3equivariant,shiba2023botan}. Additional efforts involve incorporating complementary physical parameters, including thermodynamic vibrational entropy and kinetic features \cite{yang2021machine,jung2022predicting, alkemade2023improving}. Among them, algorithms like SVM, GBDT, Linear regression, and GNN are commonly employed for building correlations in a supervised manner. Despite advancements, many approaches lack transparency, compromising interpretability. On the other hand, addressing these challenges via unsupervised machine learning, while reducing reliance on specific signals, unfortunately results in a notable decline in predictive performance. 
Therefore, finding a balance between performance and interpretability remains an unresolved challenge in glassy physics.

To address this challenge, our paper aims to improve the interpretability of machine learning models while maintaining predictive performance. We introduce a model called Structural Constraint Neural Network (SCNN) to unravel the influence of structure on dynamics in metallic glass. SCNN autonomously identifies representative particles that reflect structural distinctions among atoms with different dynamics. In contrast to other methods, the local structures of these representatives act as constraints for dynamic prediction. Then, SCNN predicts dynamics based on the similarity between each particle and the identified structural constraints. Our approach provides a transparent and easily interpretable method, offering a novel perspective on the structure-dynamic correlation in metallic glass.

\section{Methods}

\subsection{Thermal activation energy data}

In this paper, we test the interpretability of our model by predicting the thermally activated $\beta$ dynamics from the structure of Cu- Zr MG samples. Specifically, the activation energy $E_{act}$ from the thermal excitations serves as the dynamic label to distinguish the liquid-like particles and solid-like particles. From the perspective of the potential energy landscape, activation energy refers to the energy required for atoms to overcome a certain barrier and become active due to the influence of temperature. The activation energy serves as a key dynamics parameter in describing the structural origin of $\beta$ dynamics of amorphous systems \cite{yang2021machine, xu2023identification}. Interpreting and predicting atomic activation energy from the structure provides practical physical insights and scientific value. 

\subsubsection{Dataset description}
To enhance the reproducibility and benchmark our method interpretability, we utilize the activation energy data from Ref. \cite{wang2020predicting}. This dataset contains six Cu-Zr metallic glass models with different cooling rates and compositions. The molecular dynamics (MD) simulations based on Lammps are used to prepare the metallic glass (MG) samples. The activation-relaxation technique nouveau (ARTn) \cite{rodney2009distribution} is applied to explore the single-atom activation energies. 
The statistical values of this data set are illustrated in the Table.\ref{tbl1}. More details regarding the production of alloy samples and the calculation of activation energy can be obtained from the referenced literature.
\subsubsection{Dataset split}
This combined data set is split for identifying high $E_{act}$ particles. We set the highest $20\%$ particles as the positive class particles (label 1), and the remaining $80\%$ as the negative class particles (label 0). Our final data set for the machine learning model consists of all positive instances along with an equal number of randomly selected negative instances, totaling 24,000 particles. Subsequently, 12,000 particles are used for training, while testing and validation each employ 6,000 atoms. After training the model, we test our model prediction performance on the test data. Next, we analysis the model results on all particles from the combined dataset. 

\begin{table}[width=.9\linewidth,cols=3,pos=h]
\caption{Data statistics of six metallic glass samples. "$5000 \times 2$" indicates a total of 2 samples, with each sample comprising 5000 atoms.}\label{tbl1}
\begin{tabular*}{\tblwidth}{@{} ccc@{} }
\toprule
Models& Cooling rate &  Number of atoms\\
\midrule
Cu50Zr50 & $10^{10}$ k/s & $5000 \times 2$  \\
Cu80Zr20 & $10^{10}$ k/s & $5000 \times 2$  \\
Cu64Zr36 & $10^9$ k/s & $10000 \times 1$ \\
Cu64Zr36 &  $10^{10}$ k/s & $10000 \times 1$   \\
Cu64Zr36 & $10^{11}$ k/s& $10000 \times 1$  \\
Cu64Zr36 & $10^{12}$ k/s& $10000 \times 1$  \\
\bottomrule
\end{tabular*}
\end{table}

\subsection{Structural Descriptors}

For activation energy fitting, we input our interpretable model with hand-crafted structural order parameters for each atom. These structural parameters are rotationally invariant features. They are designed to measure the local density and orientation order with n-fold symmetry. 
\subsubsection{Radial density}
Local density is assessed through radial density functions. Previous studies usually use the Gaussian radial basis function, which assesses the density at distance $r$ with thickness $2\delta$ from the center atom. The main drawback of this function is its excessive feature redundancy, leading to an overly long feature dimension. This implies that the model requires a greater number of fitting parameters. Here we propose to use the Bessel basis function, which has shown a comparable performance in the machine learning community. These descriptors are defined as:
\begin{equation}
        B_{i}^{(0)}(r_c,n, t) = \sum_{j\in \mathcal{N}_{i}, t_j = t} \sqrt{\frac{2}{r_c}}\frac{sin(\frac{n\pi}{r_c}r_{ij})}{r_{ij}}
\end{equation}
with $i$ is the central atoms, $t$ is the atom type, and $r_ij$ is the distance between atom $i$ and $j$, $r_c$ is cutoff distance, and $n$ is the $n$-th of basis function. The atom type is Cu or Zr in this study. The cutoff distance $r_c$ is set as 5.0\AA. We only choose the 8-basis function for encoding the distance. Then, the summation separately goes over the selected particles $j$ in the system belonging to particle species $t$. Thus, the total radial parameters for particle $i$ have a length of 16, with each species having 8 parameters. It can be observed that there are fewer features compared to radial basis functions.

Note that instead of using a longer range cutoff to include the neighbor particle $j$, we concentrate on the radial distribution within a short range $r_c=$ 5.0\AA, and thus we have $\mathcal{N}_i = \{r_{ij}\leq{r_c}\}$. In the next, we will utilize shell-averaging \cite{PaperEmanuele} to obtain the long-range representation from these short-range parameters. Shell averaging means coarsening the structure parameters by averaging them over the nearest neighbors. Thus, $B_{i}^{(0)}$ denotes the original parameters of particle $i$ before performing shell-averaging.

\subsubsection{Orientation order}
Local orientation order is captured by the systematic angular order expansion of the local radial density. It can represent the angular distribution within a shell at distance $r$. The expansion is performed by first combining the radial basis function with the spherical harmonics and then summing over the neighbors. Formally, it is expressed as :
\begin{equation}
    q_i^{(0)}(l,m.r,n) = \frac{1}{C}\sum_{j\in \mathcal{N}_{i}} \sqrt{\frac{2}{r_c}}\frac{sin(\frac{n\pi}{r_c}r_{ij})}{r_{ij}}Y_l^m(\mathbf{r}_{ij})
\end{equation}
where $Y_l^m$ is the $l$-th spherical harmonic function of relative position $\mathbf{r}_{ij}$ between particle $i$ and $j$, and m ranges from $-l$ to $l$. Here, we still use the Bessel function to encode the radial distribution. The aggregation is then normalized by the factor
\begin{equation}
    C = \sum_{j\in \mathcal{N}_{i}} \sqrt{\frac{2}{r_c}}\frac{sin(\frac{n\pi}{r_c}r_{ij})}{r_{ij}}.
\end{equation}
The rotationally equivariant parameters $q^{(0)}$ are transformed to the invariant parameters by calculating the norm of edge degree $l$. Specifically, the norm of $q_i^{(0)}$ is defined as:
\begin{equation}
    Q_i^{(0)}(l,r,n) =\sqrt{\frac{4\pi}{2l+1}\sum_{m = -l}^{m = l}|q_i^{(0)}(l,m,r,n)|^2}. 
\end{equation}
Similar to the bond-order parameters, the parameters of degree $l$ capture the $l$-fold symmetry of the local structure within the $n$-th shell.

Here, both local density and local angular order are short-range parameters. To obtain a more comprehensive representation in the long-range, we perform shell averaging on these two parameters. Taking the local radial density parameter as an example, the result of the $k$-th shell averaging can be expressed as:
\begin{equation}
\mathbf{B}_{i}^{(k)} = \frac{1}{\sum_{{j}\in{\mathcal{N}_{j}}}e^{-r_{ij}/r_c}}\sum_{{j}\in\mathcal{N}_{j}}e^{-r_{ij}/r_c}\mathbf{B}_j^{(k-1)}
\end{equation}
where $\mathbf{B}_{i}^{(k)}$ denotes the results of $k$-th shell averaging of the previous state $\mathbf{B}_{i}^{(k-1)} $.

The complete descriptors for particle $i$ consist of a combination of structural parameters from each shell up to $k=2$. The final feature for particle $i$ is formulated as:
\begin{equation}
    \mathbf{X}_{i} = \left[\mathbf{B}_i^{(0)}\oplus\mathbf{B}_i^{(1)}\oplus\mathbf{B}_i^{(2)}\oplus\mathbf{Q}_i^{(0)}\oplus\mathbf{Q}_i^{(1)}\oplus\mathbf{Q}_i^{(2)}\right],
\end{equation}
where $\oplus$ denotes the concatenation. Thus, a total of 264 structural parameters are assigned to each particle. Before feeding these parameters into the machine learning model, they are first standardized for stable training.

\begin{figure*}
	\centering
		\includegraphics[]{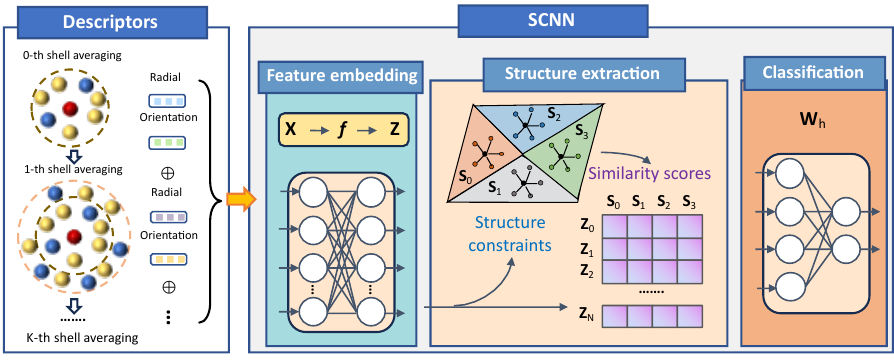}
	\caption{The architecture of structure constraint neural network (SCNN). Physical descriptors model the local radial density and bond orientation for each atom. SCNN takes these descriptors as the input, learns structural constraints, and produces classification results.  }
	\label{FIG:1}
\end{figure*}

\subsection{SCNN}
In this section, we present our interpretable deep model for activation energy prediction. The interpretability is achieved by the introduction of specific architectural biases in the model. Our main idea is from the Prototypical Neural Network (PNN) in image recognition \cite{snell2017prototypical}. The PNN is a supervised learning algorithm designed for classification tasks, especially in scenarios with limited samples per category. PNN aims to learn the prototypes for each class. Specifically, the prototypes serve as the central point for the features of samples in that category, representing the common features for each class. 
For a given new input, we can obtain the category probabilities by computing the distance between the input sample and the prototypes of each class. Overall, it provides an effective approach for the model interpretability by learning the basic concept of prototypes \cite{chen2019looks, wang2021interpretable, ragno2022prototype}. Since these prototypes can constrict the structural representation space, we refer to them as structural constraints throughout this paper. Next, we present our Structural Constraint Neural Network (SCNN).
\subsubsection{Architecture}
As shown in Fig.{}, the architecture of SCNN only consists of a feature embedding layer $f$, followed by a structure extraction layer $g$, and a final classifier layer $h$ without bias.

Given the structural parameters $\mathbf{X}_i$ for particle $i$, the feature embedding layer uses a Multilayer Perceptron (MLP) with one hidden layer to reduce the input from 264 dimensions to 32 dimensions. The structure extraction layer learns the structure constraints and calculates the similarity between all particles and the structure constraints. Note that representing high $E_{act}$ or low $E_{act}$ particles using a single structure may lack expressiveness due to the underlying complex structure. In this paper, we learn 2 structural constraints for each class. Thus, there are consists of 4 structural constraints and each has a dimension of 32. A structural constraint is denoted as $\mathbf{S}_v$ with $v\in\{0, 1,2,3\}$ indicates $v$-th structural constraint. Finally, these constraints are organized into a matrix:
\begin{equation}
\mathbf{S} = [\underbrace{\mathbf{S}_0, \mathbf{S}_1}_{\mathbf{S}^{(0)}},\underbrace{\mathbf{S}_2,\mathbf{S}_3}_{\mathbf{S}^{(1)}}]\in\mathbb{R}^{32\times{4}},\end{equation}
where the first two columns are the structural constraints for low $E_{act}$, and the last two columns are for high $E_{act}$.
The structure extraction layer then computes the cosine distances between $\mathbf{S}$  and all embeddings. The distances are then viewed as the similarity scores, which are taken as input for the final classification layer with a linear layer. The output logistic regression values are then used for classification.

Formally, these three layers can be summarized as follows:
\begin{eqnarray} 
    \mathbf{Z}_i &=& \sigma(\mathbf{W}_g \cdot (\sigma(\mathbf{W}_f \cdot \mathbf{X}_i + \mathbf{b}_f) + \mathbf{b}_g)) \\
    \mathbf{D}_i &=& \cos(\mathbf{Z}_i,  \mathbf{S}) \\
    \hat{c}_i &=& h(\mathbf{W}_h \cdot \mathbf{D}_i)
\end{eqnarray}
Here, $\mathbf{W}_f, \mathbf{W}_g, \mathbf{W}_h $ are learnable parameters, $\sigma$ is the activation function, $\mathbf{D}_i\in{\mathbb{R}^4}$ is the cosine similarity between the particle embedding $\mathbf{Z}_i$ and the structure constraints $\mathbf{S}$, and $\hat{c}_i$ is the predicted label of particle $i$.
\subsubsection{Training}

In this paper, we aim to learn a physically meaningful $\mathbf{S}$ from all training particles. To achieve this, the training approach of SCNN differs from traditional neural networks. In summary, the training of SCNN consists of four stages:
(1) initialization; (2) training of all layers except the last classification layer; (3) update of structural constraints; (4) optimization of the classification layer.

\textbf{Step 1} The structure constraints are first initialized randomly. Next, the weight matrix $\mathbf{W}_h\in\mathbb{R}^{2\times4}$ in the last classification layer is initialized as the following:
\begin{equation}
    \mathbf{W}_h = \begin{bmatrix}
    1.0,& 1,0,& -0.5,& -0.5\\
    -0.5,&-0.5,&1.0,&1.0
    \end{bmatrix}.
\end{equation}
$mathbf{W}_h$ connects the cosine similarity and the logits of class. In this scenario, the model promotes an enhancement in similarity between a particle and its corresponding structural constraints by increasing the predicted probability for the correct class based on the supervisory signal.

\textbf{Step 2} During this stage, the optimization of the model with the final classification layer is frozen. The loss function is defined as:
\begin{equation}
    \mathcal{L} = \mathcal{L}_{\text{Class}} +  \lambda_1 \mathcal{L}_{\text{Clst}} +  \lambda_2 \mathcal{L}_{\text{Sep}} +  \lambda_3 \mathcal{L}_{\text{SC-Orth}} +  \lambda_4 \mathcal{L}_{\text{SC-Sep}},
\end{equation}
where
\begin{eqnarray}
    \mathcal{L}_{\text{Class}} &=& \frac{1}{N}\sum_{i = 1}^{N} \text{CrossEntropy}(\hat{c}_i, c_i) \\
    \mathcal{L}_{\text{Clst}} &=& \frac{1}{N}\sum_{i = 1}^{N} \min_{v:\mathbf{S}_v^{c=c_i}} \left( -\frac{<\mathbf{Z}_i, \mathbf{S}_v^{c=c_i}>}{||\mathbf{Z}_i||} \right) \\
    \mathcal{L}_{\text{Sep}} &=& \frac{1}{N}\sum_{i = 1}^{N} \min_{v:\mathbf{S}_v^{c\neq{c_i}}} \left( \frac{<\mathbf{Z}_i, \mathbf{S}_v^{c\neq{c_i}}>}{||\mathbf{Z}_i||} \right) \\
    \mathcal{L}_{\text{S-Orth}} &=& \sum_{c = 0}^1 ||{\mathbf{S}^{(c)}}^T\cdot{{\mathbf{S}^{(c)}}}-\mathbf{I}||^2_F \\
    \mathcal{L}_{\text{S-Sep}} &=& -\frac{1}{\sqrt{2}}||{\mathbf{S}^{(0)}}\cdot{{\mathbf{S}^{(0)}}^T}-{\mathbf{S}^{(1)}}\cdot{{\mathbf{S}^{(1)}}^T}||^2_F
\end{eqnarray}
Here, $\lambda_1$, $\lambda_2$, $\lambda_3$, and $\lambda_4$ are hyper-parameters to balance these lost functions. 

In these loss functions, $\mathcal{L}_{\text{Class}}$ is the cross-entropy loss, where $N$ is the number of particles. $  \mathcal{L}_{\text{Clst}}$ encourage the structural embeddings close to the structural constraints of the same class, while $\mathcal{L}_{\text{Sep}}$ encourage the structural embeddings far away from the structural constraints of the different class. $<\cdot, \cdot>$ denotes the dot product. To learn meaningful structural constraints, we also applied the last two loss functions acting on the structural constraints \cite{wang2021interpretable}. The orthogonality loss function $\mathcal{L}_{\text{S-Orth}}$ ensures that the two constraints of the same class are not too close, thereby ensuring that we can learn two different structures for $E_{act}$ or low $E_{act}$. $||\cdot||^2_F$ stands for the Frobenius norm of the matrix.
Meanwhile, the structural separation loss function $\mathcal{L}_{\text{S-Sep}}$ further ensures the dissimilarity of structure between different classes, ensuring that the structural constraints for high $E_{act}$ differ from those for low  $E_{act}$. After training with a fixed number of epochs in this stage, we move on to the next phase.

\textbf{Step 3}
This step is the key to learning meaningful structural constraints.
To enhance the transparency and traceability of structural constraints, we have implemented a replacement strategy. Specifically, we select particle embeddings most similar to the structural constraints from the training set and replace the original structural constraints with them. This step aligns our structural constraints with specific particles, thereby further strengthening interpretability.

Formally, the replacement of structure constraint $\mathbf{S}^{(c)}_{v}$ can be described as:
\begin{equation}
\mathbf{S}_{v} = \mathbf{Z}_{\tilde{i}}, \text{ where }
    \tilde{i} = \underset{i\in{\mathcal{I}}}{\text{argmax}} <\mathbf{Z}_i, \mathbf{S}_{v}>.
\end{equation}
The set $\mathcal{I}$ contains all training particles.

\textbf{Step 4} The parameters of the first two layers are frozen, and the parameter $\mathbf{W}_h$ of the classification layer is optimized by minimizing the following loss function:
\begin{equation}
    \mathcal{L}_h = \mathcal{L}_{\text{Class}} + \lambda_5\mathcal{R}(\mathbf{W}_h),\label{eq19}
\end{equation}
where
\begin{equation}
    \mathcal{R}(\mathbf{W}_h)=\left|\mathbf{W}_h^{(0,2)}\right|+\left|\mathbf{W}_h^{(0,3)}\right|+\left|\mathbf{W}_h^{(1,0)}\right|+\left|\mathbf{W}_h^{(1,1)}\right|.
\end{equation}
$\left|\mathbf{W}_h^{(k,j)}\right|$ is the $l_1$ norm of the $(k,j)$-th entry of $\mathbf{W}_h$.  In addition to the classification loss, there is also a sparse constraint on the weight matrix in equation \ref{eq19}. The hyper-parameter $\lambda_5$ balances the importance of each term. For the initially fixed values of -0.5, we aim to force them towards a value of 0. This setting ensures that classification relies as much as possible on the structure constraints of the positive class and less on the constraints of negative classes.

By repeating steps \textbf{2} to \textbf{4} until convergence, we can obtain the final predictions of particle dynamics and capture meaningful structural constraints for different classes.

\begin{figure*}
	\centering
		\includegraphics[]{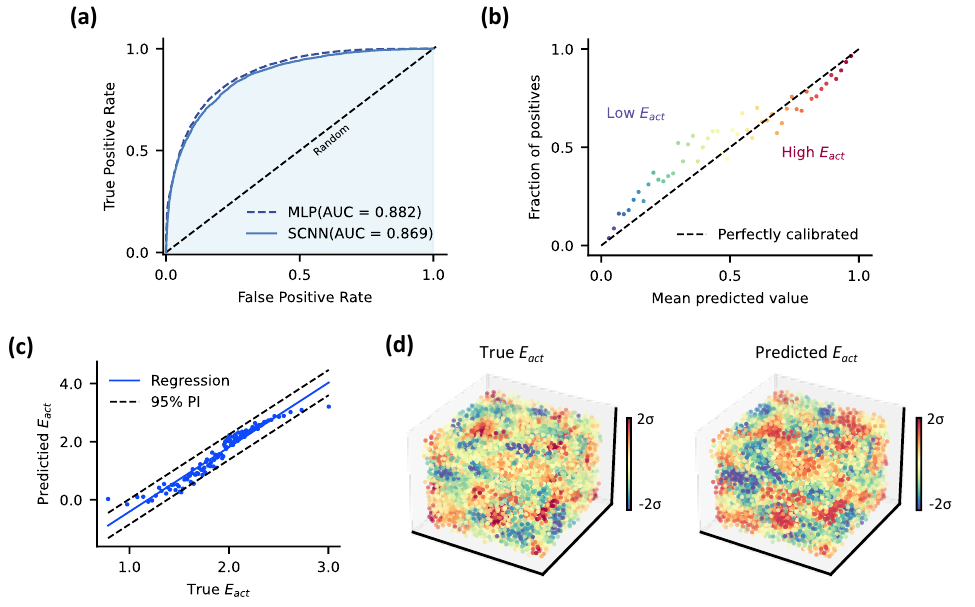}
	\caption{\textbf{(a)} The ROC curves for classifying High $E_{act}$ atoms using MLP and SCNN. TPR denotes the true positive rate and FPR denotes the false positive rate. \textbf{(b)} Calibration Curve for evaluating the reliability of probabilistic output.\textbf{(c)} Coarse-grained scatter plot between the true $E_{act}$ and predict $E_{act}$.  Each point is obtained by coarse granulation of about 100 atoms. \textbf{(d)} Visualization of true $E_{act}$ and predict $E_{act}$ in the Cu64Zr36-$10^9$ K/s metallic glass sample. In particular, the range of the color map is centered on the mean plus or minus two standard deviations ($\sigma$) of the data.}
	\label{FIG:2}
\end{figure*}

\section{Results}

\subsection{Predictive power of SCNN}
We train our SCNN model on the training data set and save the best model on the validated data set. Subsequently, we assessed the prediction performance of our SCNN model on the test dataset using the area under the receiver operating characteristic curve (AUC-ROC) as the scoring metric.

The area under the receiver operating characteristic curve (AUC-ROC) serves as the scoring metric. The ROC curve effectively illustrates the trade-off between correctly identifying positive atoms (true positives) and erroneously classifying negative atoms as positive (false positives) at different thresholds. A perfect classification is represented by an AUC-ROC value of 1.0, while 0.5 suggests random chance.
To make a comparison, we also train a Multi-Layer Perceptron (MLP) with one hidden layer to directly classify high $E_{act}$ atoms using raw physical descriptors. Figure \ref{FIG:2}(a) depicts the ROC curves for classifying atoms with high $E_{act}$. The MLP model achieves an AUC-ROC of 0.882. Remarkably, the SCNN model, relying solely on the 4-dimensional similarity scores, exhibits no significant performance drop and also attains a high AUC-ROC of 0.869. This underscores the robust capability of the SCNN model in distinguishing the structures associated with high $E_{act}$ atoms from the remaining structures with low $E_{act}$.

Figure \ref{FIG:2}(b) displays the Calibration Curve, which assesses the reliability of the probabilistic outputs of SCNN. Notably, the fraction of high $E_{act}$ atoms is strongly influenced by the output probability values. For atoms with predicted values below 0.5, the proportion of true values corresponding to high $E_{act}$ atoms is notably low. As the predicted values increase, the proportion of true values for high $E_{act}$ atoms gradually rises. This result further underscores the exceptional performance of SCNN.

To further investigate the correlation between model predictions and activation energy, Figure \ref{FIG:2}(c) presents a scatter plot depicting the relationship between model logits and true $E_{act}$. Each point on the plot represents the mean value of approximately 100 atoms. The distribution between predicted values and actual values appears to follow a linear relationship. Subsequently, in Figure \ref{FIG:2}(d), we color atoms in the Cu64Zr36-$10^9$ K/s metallic glass based on their true and predicted $E_{act}$. Here, we can observe that regions with high $E_{act}$ correspond to elevated values of output logits, while low $E_{act}$ align with lower prediction values. 
\begin{figure*}[t]
	\centering
		\includegraphics[]{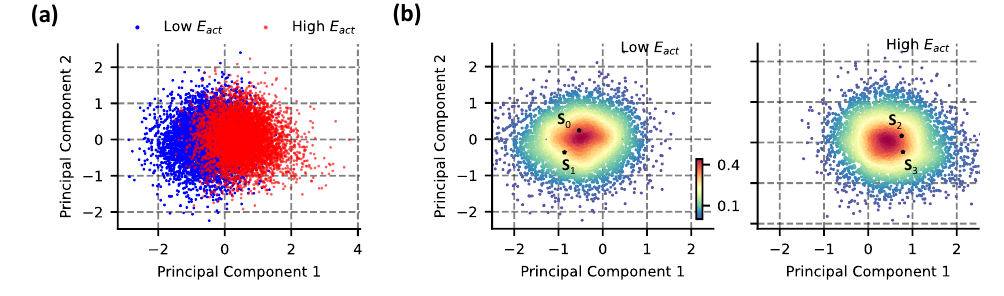}
	\caption{\textbf{(a)} Principal component analysis (PCA) to reduce the particle embeddings to two dimensions. Atoms of high $E_{act}$ in the test data set are colored red, while atoms of low $E_{act}$ are colored red.    \textbf{(b)} The density distribution of the reduced structural embeddings for low (left) and high (right) $E_{act}$ atoms. The black points indicate the position of structure constraints.}
	\label{FIG:3}
\end{figure*}
\begin{figure}[t]
	\centering
		\includegraphics[]{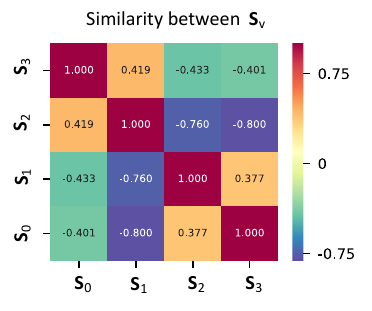}
	\caption{Heatmap of the cosine similarity between the four structure constraints. The range of cosine similarity is $[-1,1]$. A higher value represents a greater similarity between features..  }
	\label{FIG:4}
\end{figure}
\subsection{Extracting structure constraints}
The above results underscore the impeccable performance of SCNN. It is essential to note that the ultimate predicted values hinge on the cosine similarity scores between particle embeddings and structure constraints. These structure constraints serve as a crucial link between local structure and activation energy. Before delving into a deeper investigation of these structures, we further validate their reasonability. 

Initially, Principal component analysis (PCA) is employed to reduce the dimensionality of particle feature embeddings into two dimensions. In Figure \ref{FIG:3}(a), the scatter plot of the reduced features in the two-dimensional space is presented. Due to the model's strong classification performance, atoms with high and low $E_{act}$ in the reduced feature space are well-clustered.

Moreover, we visualize the density of reduced features separately for high and low $E_{act}$ atoms in Figure \ref{FIG:3}(b). High values indicate that the features of particles are densely distributed, representing common characteristics of atoms. The positions of the structural constraints are also indicated. It can be observed that the structural constraints used to characterize high and low-activation energy atoms are close to the centers of their respective clusters. This suggests that these constraints encompass the common structural features exhibited by atoms. Additionally, the constraints within each category are well-separated, with no overlap, indicating that they represent distinct structures. Thus, through the training strategy outlined in section 2.3.2, we have successfully derived meaningful structural constraints from SCNN.

Additionally, the cosine similarity analysis between these structure constraints further substantiates these findings. As illustrated in Figure \ref{FIG:4}, there is notable similarity among constraints within the same class, while constraints from distinct classes display significant differences. 

\begin{figure}[t]
	\centering
		\includegraphics[]{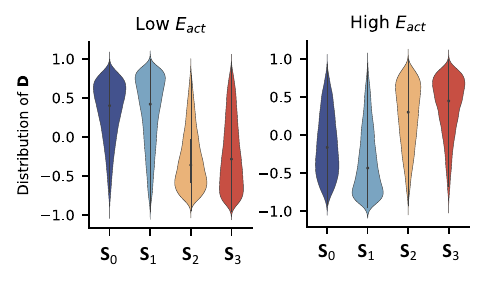}
	\caption{The distribution of the cosine similarity scores under different $E_{act}$.   }
	\label{FIG:5}
\end{figure}
\subsection{Connecting structure constraints with $E_{act}$ }

In this section, we explore the rationale behind SCNN's efficient prediction of $E_{act}$ based on the similarity scores. Figure \ref{FIG:5} illustrates the distribution of similarity scores at different $E_{act}$. For low $E_{act}$, the scores corresponding to the $\mathbf{S}_0$ and $\mathbf{S}_1$ structural constraints are concentrated around 0.7, while those for the $\mathbf{S}2$ and $\mathbf{S}3$ structural constraints center around -0.7. This indicates that the structural features of low $E_{act}$ resemble the first two structural constraints and differ significantly from the latter two. Conversely, in high activation energy, the situation is reversed. Thus, SCNN can leverage this distribution to easily differentiate between high and low $E{act}$ atoms based on similarity scores.

\begin{figure*}[t]
	\centering
		\includegraphics[]{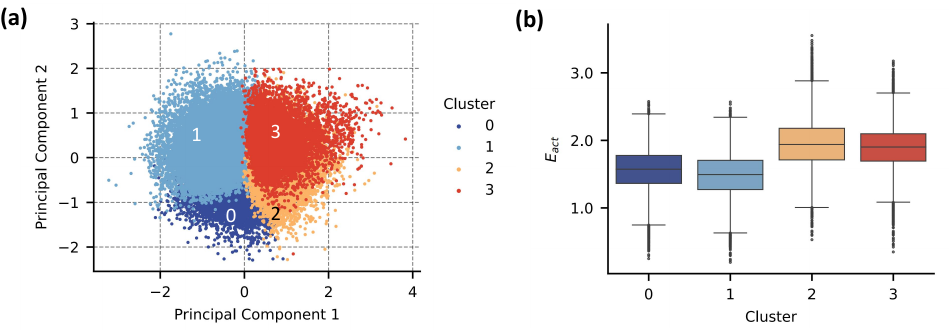}
	\caption{\textbf{(a)} PCA reduction of the feature embeddings for all atoms in the combined dataset. Each atom is colored by its cluster label. \textbf{(b)} Distribution of the true $E_{act}$ in each cluster.    }
	\label{FIG:6}
\end{figure*}
\begin{figure*}[h]
	\centering
		\includegraphics[]{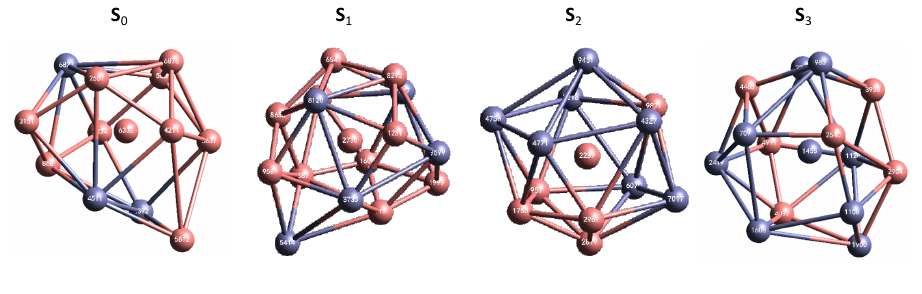}
	\caption{ The local structures of particles corresponding to the four structural constraints are characterized by the Largest Standard Cluster Analysis (LaSCA). The visualization of these microscopic structures is implemented using the visualization software of LaSCA \cite{tian2023lasca}.}
	\label{FIG:7}
\end{figure*}

To further validate the effectiveness of cosine similarity between atomic structures and the four structural constraints, we perform clustering on all atoms based on their similarity scores. Specifically, we identify the index corresponding to the highest similarity score for each atom and assign it as the structural label for that atom. It is crucial to note that the clustering is based on the atoms' structural characteristics, independent of their dynamics. Consequently, atoms in cluster $v$ exhibit the highest similarity to the $\mathbf{S}v$ structural constraints. 

Figure \ref{FIG:6}(a) provides a two-dimensional visualization of all atoms' embeddings after PCA reduction, with each color representing a distinct cluster label. Each cluster demonstrates a well-defined aggregation, affirming the reliability of structure classification based on similarity. Furthermore, the separation between different clusters signifies distinct structures among them. Subsequently, Figure \ref{FIG:6}(b) depicts the $E_{act}$ distribution within each cluster. Clusters 0 and 1 predominantly exhibit low $E_{act}$, while clusters 2 and 3 concentrate on high $E_{act}$. Thus, we can infer that these structural clusters reflect different dynamic properties. 

\begin{table}[width=.9\linewidth,cols=2,pos=h]
\caption{The LaSC of the structures in Figure \ref{FIG:7}. A LaSC is characterized by the number and types of Common Neighbor Sub-clusters (CNS). The notation n/Sijk is used, where n represents the number of CNS, and Sijk indicates the type of CNS.}\label{tbl2}
\begin{tabular*}{\tblwidth}{@{} ccc@{} }
\toprule
Structure& LaSC  \\
\midrule
$\mathbf{S}_0$ &  [2/S433, 2/S444, 2/S544, 4/S555, 2/S666] \\
$\mathbf{S}_1$ & [3/S444, 6/S555, 6/S666]\\
$\mathbf{S}_2$ & [12/S555] \\
$\mathbf{S}_3$ &  [1/S421, 2/S433, 4/S544, 6/S555, 1/S666]  \\
\bottomrule
\end{tabular*}
\end{table}

\subsection{Analysis of the structure constraints}
\begin{figure*}[th]
	\centering
		\includegraphics[]{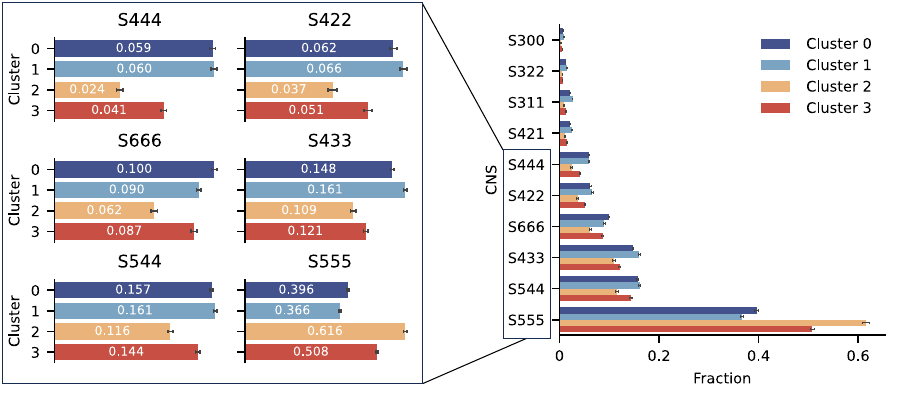}
	\caption{The fraction of top-10 CNS at each atom cluster (left) and the Enlarged view of the top-6 CNS (right). The mean value of fractions is indicated in the center of each bar.  }
	\label{FIG:8}
\end{figure*}

\begin{table}[width=0.9\linewidth,cols=3,pos=h]
\caption{The top-10 LaSCs in all CuZr samples. Each LaSC is labeled by its topologically close-packed (TCP) index. }\label{tbl3}
\begin{tabular*}{\tblwidth}{@{\extracolsep{\fill}} ccc@{} }
\toprule
LaSCs & TCP Label & Count \\
\midrule
$[12/555]$ & Z12 & 5140 \\
$[2/433, 2/544, 8/555]$ & 1-Z12 & 3324\\
$[1/444, 10/555, 2/666]$ & A13 & 1308\\
$[2/444, 8/555,2/666]$ & B12 & 1073\\
$[3/444, 6/555, 4/666]$ & C13 & 906\\
$[12/555, 4/666]$ & Z16 & 758\\
$[1/433, 3/544, 9/555, 2/666]$ & 1-A15 & 747\\
$[2/422, 2/433, 2/544, 6/555]$ & 2-Z12 & 727\\
$[2/444, 8/555, 1/666]$ & B11 & 666\\
$[1/421, 3/433, 3/544, 5/555]$ & 2-Z12 & 640\\
\bottomrule
\end{tabular*}
\end{table}
\begin{figure}[th]
	\centering
		\includegraphics[]{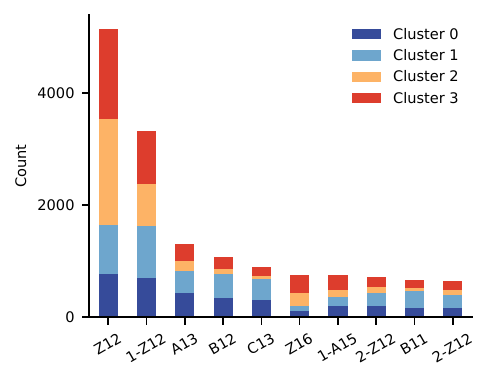}
	\caption{ The count of top-10 LaSCs at each atom cluster.  }
	\label{FIG:tcp}
\end{figure}
Our SCNN transforms the dynamic prediction problem into a similarity comparison between atomic structural features and structural constraints. Upon verification, these structural constraints can identify primary distinctions between solid-like and liquid-like structures. In SCNN, each structural constraint corresponds to the local structure of a particle. We designate these particles as representative particles. Naturally, the structural distinctions among these representative particles can elucidate the primary cause of the dynamic differences, a capability not achievable with current machine-learning methods.
Therefore, a detailed exploration of the atomic environments for these representative particles can get a deeper understanding of the structure-dynamics relationship.

We identified these representative particles and visualized their local atomic structures, as depicted in Figure \ref{FIG:7}. Here, the neighboring particles are obtained through the Largest Standard Cluster Analysis (LaSCA) \cite{tian2011new,tian2014structural}. The LaSCA algorithm decomposes the local structure of each particle into a series of Common Neighbor Sub-clusters (CNS) in a physically meaningful manner. The structural features of CNS denoted as the CNS index, can be characterized using Sijk. Here, i represents the total number of Common Near Neighbors (CNNs), j represents the total number of edges within the CNN, and k represents the length of the longest path within the CNN. For instance, the CNS index of S555 represents the local 5-fold symmetry structure with 5 common neighbors and 5 edges connecting these 5 neighbors, while the longest path of the edge is 5.

From Figure \ref{FIG:7}, we can observe that the latter two structural constraints exhibit pronounced symmetry compared to the former two. We also detail their LaSC in Table \ref{tbl2}. It can be observed that all these structures contain S555, indicating the widespread presence of 5-fold symmetry in amorphous structures. However, their abundances vary, with S555 constituting 0.33, 0.40, 0.1, and 0.43 in $\mathbf{S}_0$, $\mathbf{S}_1$, $\mathbf{S}_2$, and  $\mathbf{S}_3$, respectively. The content of S555 is higher in $\mathbf{S}_2$ and $\mathbf{S}_3$ compared to $\mathbf{S}_0$ and $\mathbf{S}_1$. Notably, $\mathbf{S}_2$ exhibits a typical icosahedral structure. Additionally, the contents of S444 and S666 are significantly higher in $\mathbf{S}_0$ and $\mathbf{S}_1$ compared to $\mathbf{S}_2$ and $\mathbf{S}_3$. In a previous study \cite{tian2017local}, it has been demonstrated that the close-packing nature of S555 is significantly superior to S444 and S666. The high-density packing contributes to the stability of the structure. Note that the $\mathbf{S}_0$ and $\mathbf{S}_1$ represent the structural features associated with liquid-like atoms, while $\mathbf{S}_2$ and $\mathbf{S}_3$ correspond to the structural features related to solid-like atoms. Therefore, the solid-like characteristics of $\mathbf{S}_2$ and $\mathbf{S}_3$ may result from the high-density packing of the structure, while the liquid-like characteristics of $\mathbf{S}_0$ and $\mathbf{S}_1$ originate from structural defects caused by low-density packing.

\begin{figure*}[hb]
	\centering
		\includegraphics[]{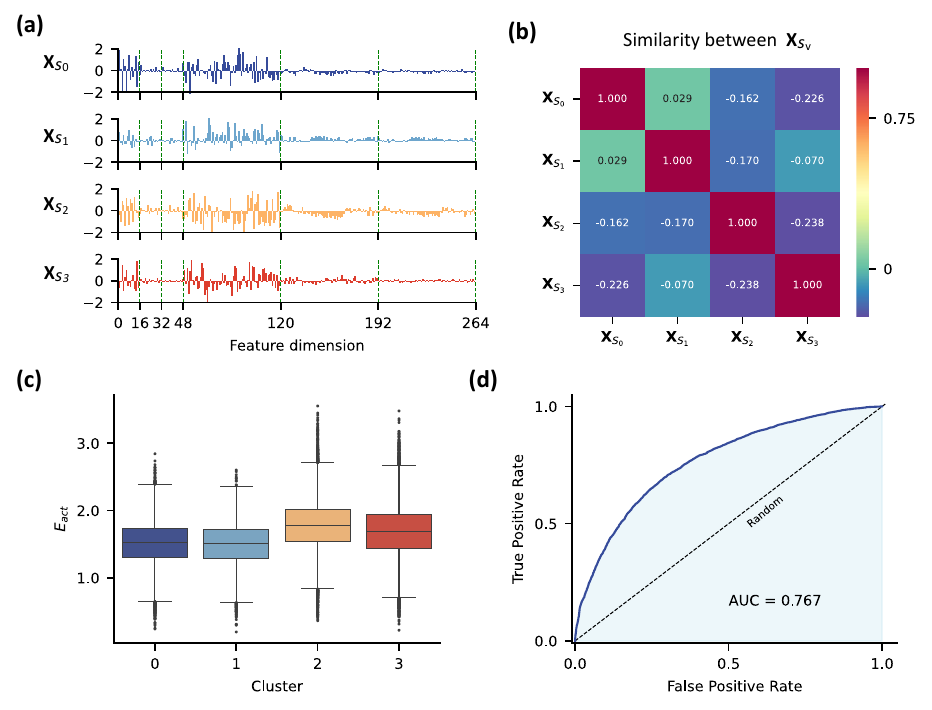}
	\caption{ \textbf{(a)} Visualization of the physical descriptors for the structure constraints. The green dashed line represents the boundary separating the radial and orientation features in different shells. \textbf{(b)} Cosine similarity between the four physical descriptors. \textbf{(c)} Distribution of true $E_{act}$ in each cluster. \textbf{(d)} ROC curve for classifying the high $E_{act}$ using the similarity scores as the atom feature.}
	\label{FIG:9}
\end{figure*}
 % [ 12/555 ]          Z12      5140 
 % [ 2/433, 2/544, 8/555 ]     1-Z12    3324
 % [ 1/444, 10/555, 2/666 ]   A13     1308
 % [ 2/444, 8/555,2/666 ]     B12    1073
 % [ 3/444, 6/555, 4/666 ]    C13      906
 % [ 12/555, 4/666 ]          Z16   758
 % [ 1/433, 3/544, 9/555, 2/666 ]  1-A15    747
 % [ 2/422, 2/433, 2/544, 6/555 ]   2-Z12   727
 % [ 2/444, 8/555, 1/666 ]     B11     666
 % [ 1/421, 3/433, 3/544, 5/555 ]    2-Z12  640

In Section 3.3, we performed clustering of atoms based on their similarity with the structural constraints. To validate the aforementioned findings, we conducted a detailed analysis of the CNS distribution within each atom cluster.

Initially, we computed the proportion of each CNS in the LaSC of each atom. Subsequently, we presented the proportion of CNS in each cluster in Figure \ref{FIG:8}, showcasing only the top 6 most abundant CNS. Apart from S555, other CNS in classes 0 and 1 exhibit significantly higher proportions than those in classes 2 and 3. These CNS may contribute to structural defects, consequently diminishing the stability of the structure. Conversely, classes 2 and 3 manifest high 5-fold symmetry structures, enhancing the stability of their structures. This reaffirms the established understanding that structural disparities primarily arise from local 5-fold symmetry \cite{hu2015five}.

Next, we further explore the distribution of LaSCs in each cluster. Table \ref{tbl3} presents the top-10 LaSCs and their number in the combined dataset. The labels for each LaSC are defined by their topologically
close-packed (TCP) characteristics, which are proposed in \cite{wu2018topologically}. For the detailed meanings of these labels, the readers can refer to \cite{wu2018topologically, zhou2021correlation}. The counts of these LaSCs at each cluster are shown in Figure \ref{FIG:tcp}. Clusters 2 and 3 contain more atoms of Z12, which represents the icosahedra structure, than clusters 0 and 1. This indicates that Z12 plays an important role in solid-like structures. Additionally, we can observe that A13, B12, and C13 are more in clusters 0 and 1, suggesting their underlying liquid-like characteristics. 
% To enhance the accuracy of the analysis results, we focus on the structural differences in these distinct clusters.
\subsection{Generalization of the structure constraints}
Finally, we conducted a study on the generalization of the four structural constraints. In the SCNN dynamic prediction process, the original input features are first mapped to an embedding space, followed by a comparison in the embedding space with the four structural constraints. The dynamics prediction is then completed based on the similarity obtained from the comparison. These steps in SCNN are designed to learn more abstract representations from the original features, improving the discrimination between liquid and solid-like structures. In other words, running the forward process of SCNN is necessary to obtain the output. Now, the question arises: can we bypass the complex computations of SCNN and directly utilize the original handcrafted physical descriptors for classification? Specifically, can we use the raw physical descriptors of the four structural constraints as the basis for dynamic prediction? If feasible, this would be highly meaningful, suggesting that these four structural constraints contain crucial information for distinguishing between liquid and solid-like structures.

Figure \ref{FIG:9}(a) illustrates the raw physical descriptors corresponding to the four structural constraints. Intuitively, these descriptors exhibit significant differences. As evident from the Consine similarity matrix in Figure \ref{FIG:9}(b), the descriptors of these four atoms show low similarity. In contrast to the results in Figure \ref{FIG:4}, these raw descriptors, without optimization or filtering by SCNN, do not reveal inter-class similarities but still preserve inter-class differences. Hence, we can still utilize these structural parameters as the basis for atomic classification. 

Subsequently, we calculated the similarity between each atom and the four physical descriptors. Based on this similarity, atoms were categorized into four classes, as done in section 3.3. The distribution of $E_{act}$ for these four classes is shown in Figure \ref{FIG:9}(b). Interestingly, we observe that the distributions in each cluster are similar to the results using the embedded feature as the cluster basis, which are shown in Figure \ref{FIG:6}(b). Despite the dynamic differences between clusters not being as pronounced as with the optimized features, it demonstrated that clustering based solely on the raw descriptors of these structure constraints still can yield a significant improvement in exploring the dynamic property. 

To further validate this, we use the 4-dimensional similarity scores computing from the physical descriptors as atom features to retrain the final classification layer of SCNN. The ROC curve on the test set is shown in Figure \ref{FIG:9}(d). At this point, the AUC value for the classification of high $E_{act}$ atoms is 0.767. This result confirms that the key structural information distinguishing between liquid-like and solid-like structures is contained within these four representative atoms. It significantly narrows down the search space for studying the correlation between structure and dynamics.

\section{Conclusion}
This paper endeavors to construct an interpretable model for comprehending the correlation between structures and dynamics in amorphous solids. Diverging from conventional machine learning models, our framework acquires representative structural constraints during training, capturing diverse structural features in dynamics. The model demonstrates its effectiveness by precisely predicting activation energies from the atomic local environment of CuZr samples.

Since each learned structure constraint corresponds to a particle, we can narrow down the vast structural search space to the local environment of these four particles. Our analysis of structural constraints identifies local 5-fold symmetry as the primary factor influencing differences in amorphous atomic dynamics. 
This finding, obtained for the first time in the field of machine learning research for amorphous solids, validates the superiority of our proposed interpretable approach. Additionally, we utilized the original descriptors of the four representative atoms to classify activation energy, achieving exceptional predictive performance. This suggests that the original physical descriptors of the four structural constraints already contain effective structural information. In the future, analyzing the medium-range structures governed by these four primary constraints, we believe, will undoubtedly contribute to understanding the structural origins of dynamic heterogeneity in amorphous alloys.

As the first attempt to introduce machine learning interpretability into the realm of amorphous materials, this undoubtedly has the potential to significantly propel research in this field. In this study, we did not compare different descriptors but rather focused on interpretability. Surprisingly, we achieved excellent predictive results using simple structural descriptors alone. Enhancing structural expressiveness, such as integrating more complex structural descriptors and designing an end-to-end interpretable framework using the graph neural network, are feasible avenues to improve both prediction accuracy and interpretability.

% \printcredits

%% Loading bibliography style file
\bibliographystyle{unsrt}
\bibliography{ref}
% Loading bibliography database
% \bibliography{cas-refs}

%\vskip3pt

\end{document}